\documentclass[aps,prl,twocolumn,groupedaddress,showpacs]{revtex4}
\usepackage{epsfig}
\usepackage[dvipsnames,usenames]{color}
\usepackage{color}
\usepackage{amsmath}

\emergencystretch=\maxdimen
\begin{document}
\title{Magnetic Correlations in a Periodic Anderson Model
with Non-Uniform Conduction Electron Coordination}
\author{N.~Hartman$^{1}$, W.-T. Chiu$^{2}$ and R.T.~Scalettar$^2$}

\affiliation{$^1$Department of Physics,
Southern Methodist University,
Dallas, Texas 75275-0175
}
\affiliation{$^2$Department of Physics, One Shields Ave.,
University of California, Davis, California 95616, USA}

\begin{abstract}
The Periodic Anderson Model (PAM) is widely studied to understand
strong correlation physics and especially the competition of
antiferromagnetism and singlet formation.
Quantum Monte Carlo (QMC) studies have focused both on
issues such as the nature of screening and
locating the quantum critical point (QCP) at zero temperature
and also on possible experimental connections to phenomena ranging from the
Cerium volume collapse to the relation of the magnetic
susceptibility and Knight shift in heavy fermions.
In this paper we extend QMC work to lattices in which
the conduction electron sites can have variable
coordination.  This situation is relevant both to recently discovered
magnetic quasicrystals and also to magnetism in doped heavy fermion
systems.
\end{abstract}

\maketitle

\section{Introduction}

The single band Hubbard
Hamiltonian\cite{rasetti91,montorsi92,gebhard97,fazekas99}
captures several of the most
fundamental consequences of electron-electron interactions in solids,
namely magnetic order and the Mott metal-insulator transition.
Although the question is still open, it may even contain
the fundamental physics of $d$-wave superconductivity in 
the cuprates\cite{white89,scalapino94}.
Multi-band Hamiltonians like the periodic Anderson model (PAM)
\cite{anderson61,doniach72}
examine what happens when two species of electrons, one
delocalized `conduction' band (often $d$), and another 
localized band (often $f$)
are present.  Here a central effect is the competition between
singlet formation\cite{kondo69}, when the conduction and localized electrons 
are strongly hybridized, and ordering of the local moments
mediated indirectly
through the Ruderman-Kittel-Kasuya-Yosida 
interaction\cite{ruderman54,kasuya56,yosida57,kittel63}.

In heavy fermion materials\cite{stewart84,stewart01}, 
this competition is believed
to explain different low temperature phases, 
e.g.~non-magnetic CeAl$_{3}$ where the $f$ moments are screened by the
conduction electrons, 
and CeAl$_{2}$ which becomes antiferromagnetic (AF) at low temperatures.
Likewise, in Ce metal itself, as well as in
other rare-earths, the tuning of the $f$-$d$ hybridization
$V_{\rm fd}$ through pressure is believed to play a crucial role in
the `volume collapse'
transition\cite{gschneider78,allen82,allen92,lavagna82,mcmahan98,held01}.

Quantum Monte Carlo (QMC) studies of the PAM have explored some of this
physics, in one\cite{blankenbecler87,fye90}, two\cite{vekic95}, 
and three dimensions\cite{huscroft99}.   The focus has
been on bipartite lattices which, at half-filling, host AF order without
frustration and are also free of the fermion
sign problem\cite{loh90,troyer05}.
QMC in infinite dimensions\cite{jarrell93}
complements work in finite $d$ by
allowing simulations at very low temperature, at or even well below the
Kondo scale, at the expense of some of the knowledge of 
correlations in space.

There is interest in understanding the magnetic correlations in 
more general geometries.  One such modification allows for intersite,
rather than on-site,
hybridization between conduction and local orbitals and
hence metallic behavior in the absence of interactions\cite{held00}.
Another motivation is provided by chemical
substitution in heavy fermion materials, either by the replacement of
some of the local moment atoms by non-magnetic ones,
as in (Ce,La)CoIn$_5$\cite{nakatsuji02}
or by changes to the conduction orbitals, as in
the alloying of Cd onto In sites in CeCo(In,Cd)$_5$ 
\cite{ParkDropletsNature2013,seo14}.
In the latter situation, $V_{\rm fd}$ is reduced locally, and AF droplets
can form around the impurity sites.
A second motivation is the recent observation of a 
quantum critical state in magnetic
quasicrystals\cite{deguchi12,watanuki12}.
In these Au-Al-Yb alloys
(Au$_{51}$Al$_{34}$Yb$_{15}$), measurements of the magnetic
susceptibility $\chi$ and specific heat $C$ diverge as $T\rightarrow 0$.
This non-Fermi liquid (NFL) behavior is associated with strongly correlated
$4f$ Yb electrons.  
These two sets of materials share a common feature which is that 
the coordination
number of the different atoms is no longer spatially uniform.
The effects of these unique local environments can be probed with
nuclear magnetic resonance\cite{curro99}.

The NFL behavior of Au-Al-Yb alloys has recently been studied by solving the 
$U=\infty$ Anderson Impurity Model (AIM) for a single local moment
coupled to conduction electrons in a quasicrystal approximant
geometry\cite{andrade15}.  The crucial result is that singular responses
in $\chi$ and $C$ occur as a consequence of 
a broad (power law) distribution of Kondo temperatures which delays
screening of a large fraction of the magnetic moments 
until very low temperatures.  

In this paper we study the PAM in two different geometries:
the Lieb lattice and 
a 2D ``Ammann-Beenker"
tiling\cite{grimm02,jagannathan07}.
Quasicrystaline approximates\cite{goldman93} for
Au$_{51}$Al$_{34}$Yb$_{15}$
are in 3D; the quasi-periodic
Ammann-Beenker tiling is a more tractable 2D alternative
for QMC, which is limited in the number of sites which
can be simulated.
Our goal is to explore the nature of magnetic correlations as a function
of $f$-$d$ hybridization, and, specifically, to understand the competition
of antiferromagnetic order and singlet formation in geometries where
the coordination number of different sites in the lattice is
non-uniform.
Our work extends that of \cite{andrade15} by examining
a dense array of local orbitals 
and also by including the effect of finite $U_{\rm f}$.
We begin with the Lieb lattice, because it 
contains two separate coordination numbers, $z=2,4$ 
while still retaining very simple lattice periodicity.  We then
turn to the more complicated quasicrystal approximant structure.
We do not at present address the anomalous NFL
behavior of the magnetic susceptibility, since those phenomena
appear to be associated only with the quasicrystal itself, and
not its approximant\cite{seo14}. 

The magnetic properties of quantum antiferromagnets in
geometries which have variable coordination number 
(the crown, dice, and CaVO lattices) have also been
studied in the context of the spin-1/2 Heisenberg Hamiltonian
\cite{wessel03,wessel05,jagannathan06}.
Surprisingly, unlike the case of regular geometries where the ordered
moment {\it increases} from the honeycomb ($z=3$)
to square ($z=4$) lattices, it is found that the local AF order parameter
{\it decreases} with $z$.  We will comment on this further in our
conclusions.

\section{Model and Methods}

\begin{figure}[h!]
\includegraphics[height=6.0cm,width=6.0cm]{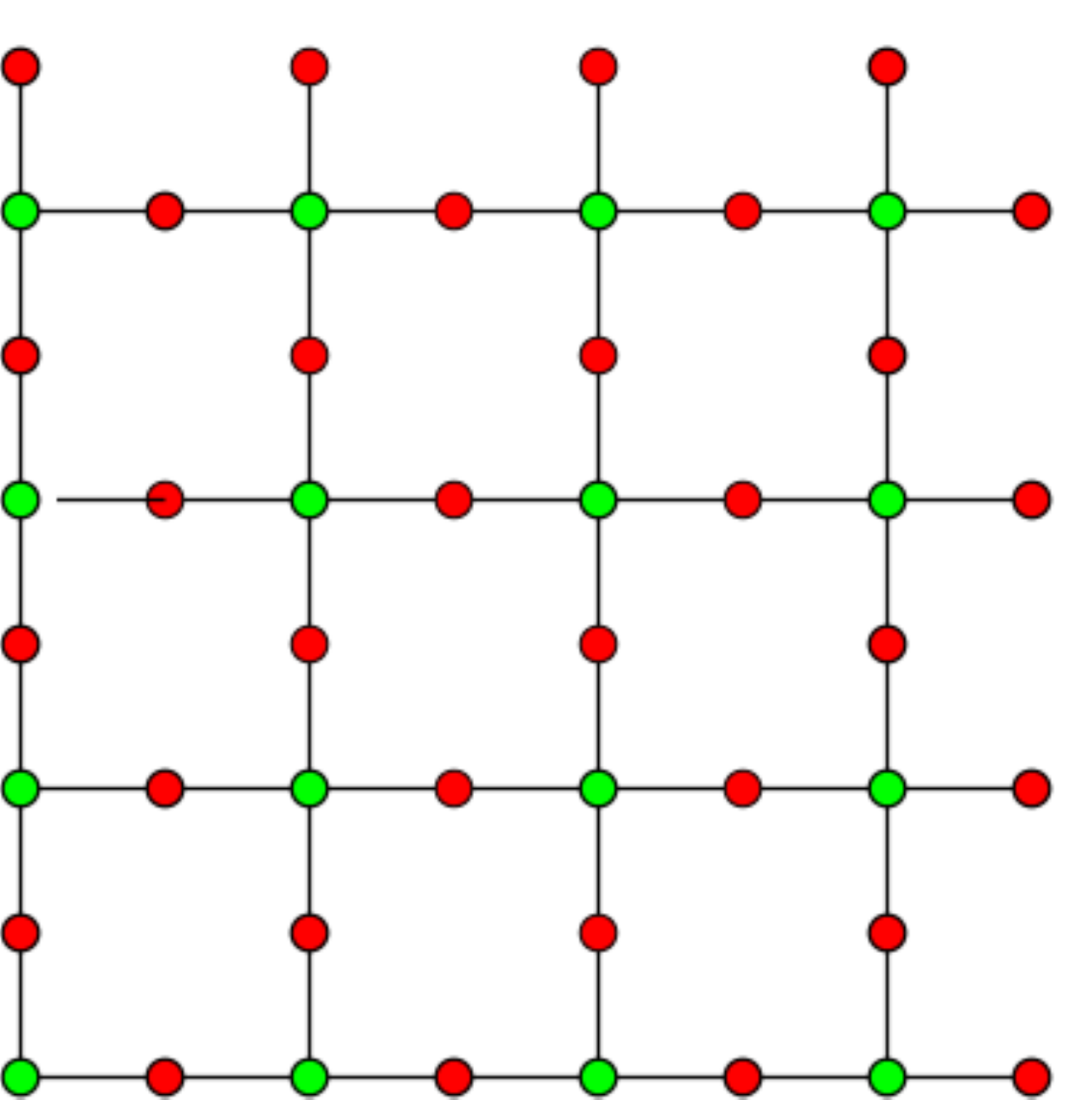}
\caption{(Color online) 
The Lieb lattice geometry under consideration in this paper.  
(Cluster shown has 4x4 unit cells with 48 sites).  
Each site contains both a conduction ($d$) orbital
and a localized ($f$) orbital, so that there is a total
of 96 sites/orbitals.  Lines correspond to the
$d$-$d$ hopping $t$, with eight possible coordinations
$z=1,2,.\cdots 8$.  We use periodic boundary conditions (pbc).
There are two possible conduction orbital coordinations, $z=2,4$.
The local $f$ orbitals are connected
to the $d$ orbital on the same site by hybridization $V_{\rm fd}$.
\label{fig:liebgeom}
}
\end{figure}

\begin{figure}[h!]
\includegraphics[height=6.0cm,width=7.0cm]{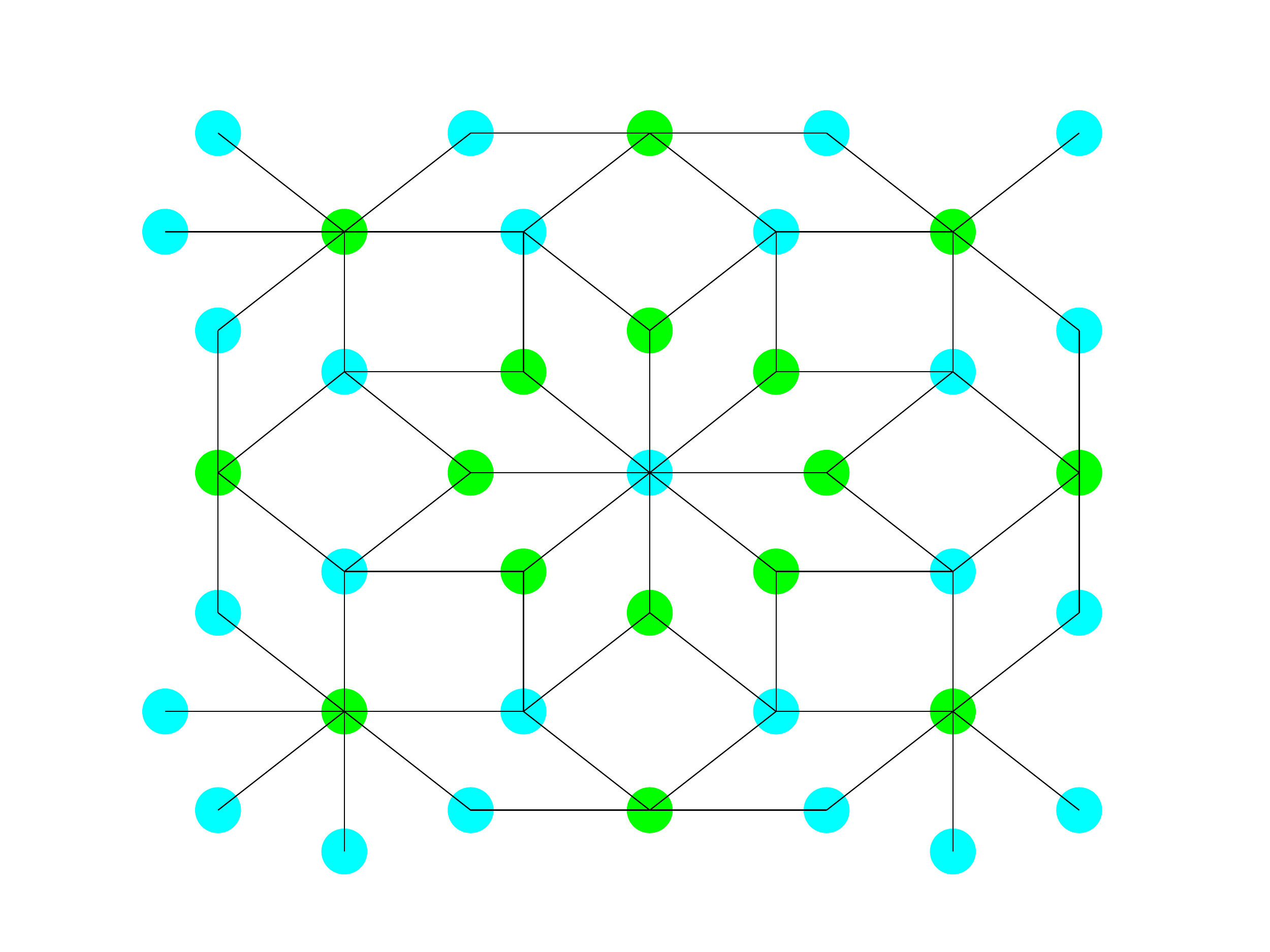} \\
\includegraphics[height=6.0cm,width=7.0cm]{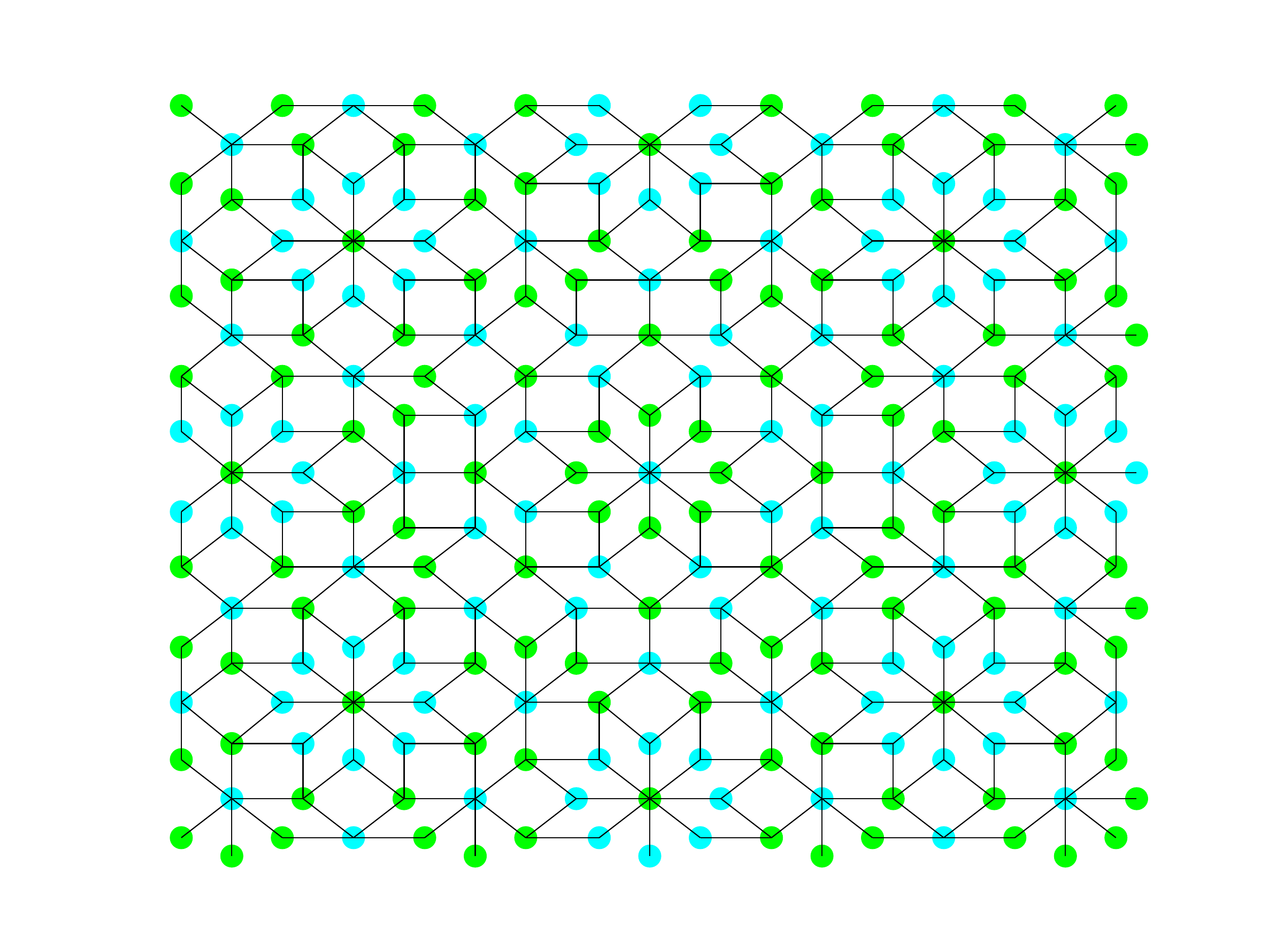}
\caption{(Color online) 
Top (bottom): Approximants to the
Au$_{51}$Al$_{34}$Yb$_{15}$ crystalline lattice for $N=41 (239)$ sites.
In each case, sites shown contain both a conduction ($d$) orbital
and a localized ($f$) orbital.  Lines correspond to the
$d$-$d$ hopping $t$.  The local $f$ orbitals are connected
to the $d$ orbital on the same site by hybridization $V_{\rm fd}$.
For this geometry we use open boundary conditions (OBC) to avoid 
frustration.  The conduction electron sites range in coordination
from $z=1$ to $z=8$.  The use of two colors
for the sites emphasizes that, despite its complexity, the geometry is still
bipartite.
\label{fig:quasigeom}
}
\end{figure}

The PAM is a tight binding Hamiltonian for which each spatial
site contains both an extended and a localized orbital,
\begin{align}
    {\cal H} = &-t \sum\limits_{\langle ij \rangle,\sigma}
(d^{\dagger}_{i\sigma}d_{j\sigma}^{\vphantom{dagger}}
+d^{\dagger}_{j\sigma}d_{i\sigma}^{\vphantom{dagger}})
        -V_{\rm fd} \sum\limits_{i\sigma}
(d^{\dagger}_{i\sigma}f_{i\sigma}^{\vphantom{dagger}}+
f^{\dagger}_{i\sigma}d_{i\sigma}^{\vphantom{dagger}})
\nonumber \\
        &+ U_{\rm f} \sum\limits_{i} (n^{f}_{i\uparrow}-\frac{1}{2})
(n^{f}_{i\downarrow}-\frac{1}{2})
\label{eq:PAM}
\end{align}
Here $t$ is the hybridization between conduction orbitals with
creation(destruction) operators 
$d_{i \sigma}^{\dagger} (d_{i \sigma}^{\phantom{\dagger}})$
on near neighbor sites $\langle ij \rangle$.
In this paper we consider the two conduction electron geometries shown in 
Figs.~\ref{fig:liebgeom},\ref{fig:quasigeom} corresponding to 
``Lieb" and ``quasicrystal" lattices respectively.
Each site of these lattices also contains a localized orbital,
creation(destruction) operators 
$f_{i \sigma}^{\dagger} (f_{i \sigma}^{\phantom{\dagger}})$.
$U_{\rm f}$ is the on-site interaction
between spin up and spin down electrons on the localized orbital, and
$V_{\rm fd}$ is the conduction-localized
orbital hybridization.
Both geometries of Figs.~\ref{fig:liebgeom},\ref{fig:quasigeom}
are bipartite.  
In ${\cal H}$ we have written the interaction term, 
in `particle-hole' symmetric form, and set the site energy difference
between f and d orbitals to zero, so that the
lattice is half-filled for all temperatures $T$ and 
Hamiltonian parameters $t,U_{\rm f},V_{\rm fd}$.
Half-filling optimizes the tendency for AF correlations, and
also allows DQMC simulations at low temperature since the
sign problem\cite{loh90} is absent.

\begin{figure}
\psfig{figure=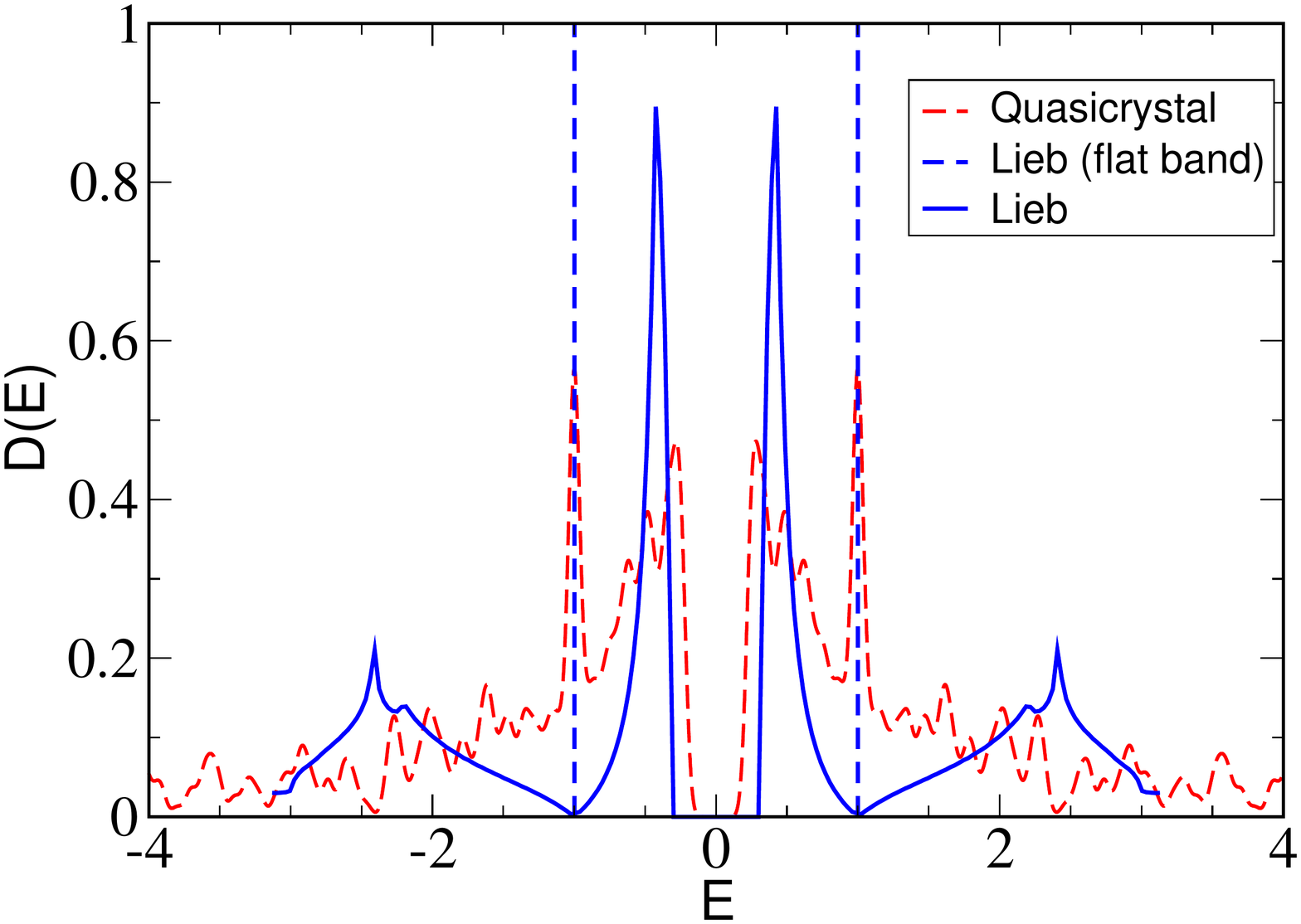,height=8.0cm,width=7.6cm,angle=0,clip} \\
\vskip-0.35in
\hskip0.1in
\psfig{figure=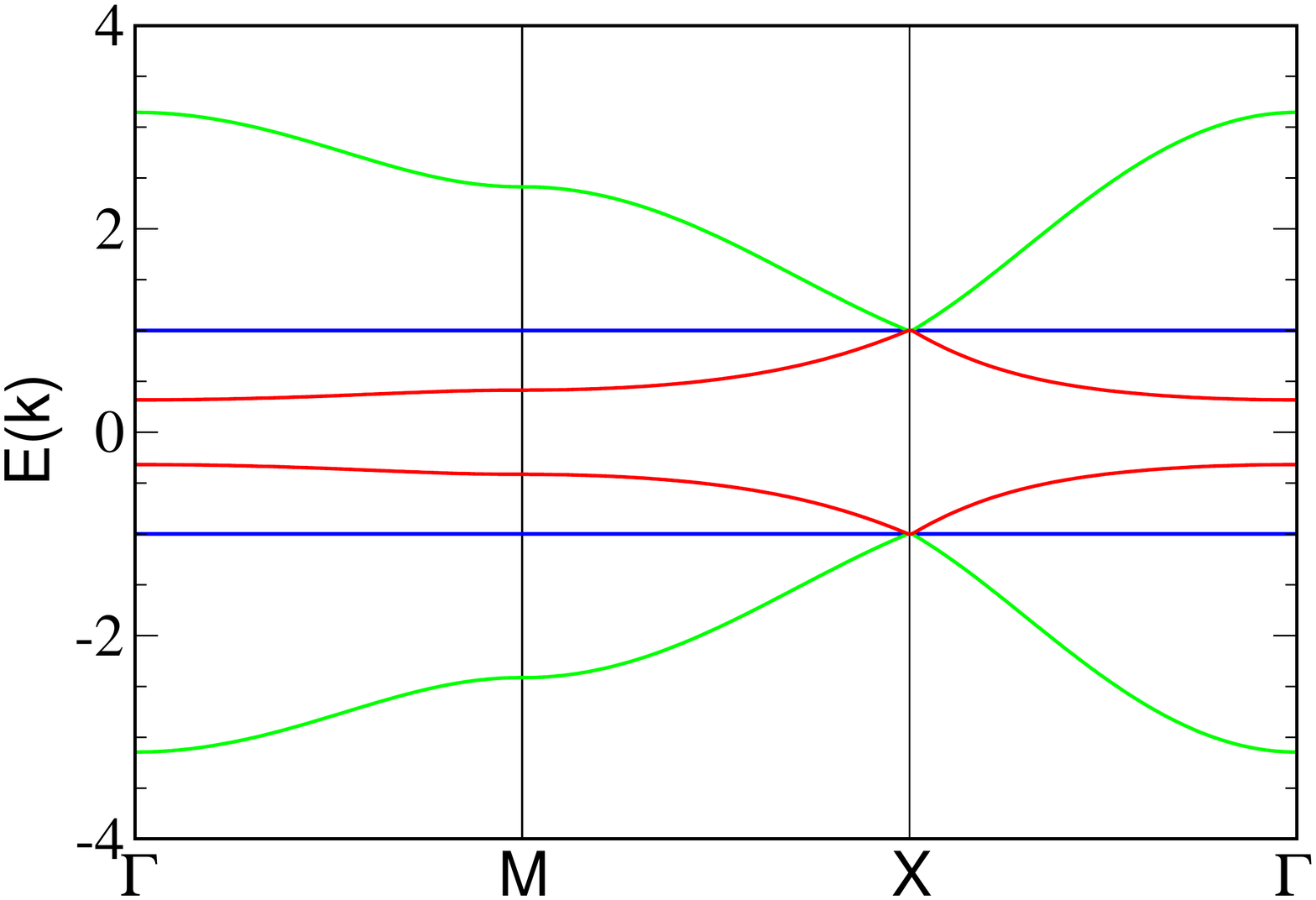,height=8.0cm,width=7.6cm,angle=0,clip} \\
\caption{(Color online) 
DOS (top) and band structure (bottom) of the PAM on a Lieb lattice.  
Here $t=1$ and $V_{\rm fd}=1$.
The two completely flat bands at $E=\pm 1$ give rise to 
$\delta$ function spikes in $D(E)$ 
which are indicated by dashed vertical lines).
$D(E=0)$ vanishes:
the system is a band insulator
at half-filling.  
The DOS for the $N=239$ quasicrystal approximant is also shown in the
top panel.  
As is the case for the Lieb lattice PAM,
the quasicrystal PAM also has a hybridization gap at $E=0$.  
The single band case is metallic\cite{andrade15,jagannathan06}.
\label{fig:Liebdos}
}
\end{figure}

Figure \ref{fig:Liebdos} shows the density of states and band structure
of the PAM on a Lieb lattice for $t=1$ and $V_{\rm fd}=1$.  
There are six bands corresponding to the
six sites (three conduction and three localized) per unit cell.  The
lattice is bipartite with four of the six sites on one sublattice
and two on the other. Hence, in accordance with Lieb's
theorem\cite{lieb89} there are two flat bands (at $E=\pm 1$).
As in the case of the PAM on a square lattice with on-site
hybridization, the half-filled lattice is a band insulator in the 
non-interacting limit.  However, by comparing calculations for
on-site and intersite $V_{\rm fd}$, the latter being
metallic at half-filling, it has been shown that many properties
of the PAM when $U_{\rm f}/t \gtrsim 4$ are insensitive to
the presence of a $U_{\rm f}/t=0$ band gap\cite{held00}.

The magnetic properties of the PAM
are characterized by intra- and inter-orbital spin-spin correlations,
\begin{align}
c^{zz'}_{\rm ff}(r) &= 
\langle f^{\dagger}_{i+r \downarrow} f^{\phantom{\dagger}}_{i+r \uparrow}
f^{\dagger}_{i \uparrow} f^{\phantom{\dagger}}_{i \downarrow} \rangle
\nonumber
\\
c^{zz'}_{\rm dd}(r) &= 
\langle d^{\dagger}_{i+r \downarrow} d^{\phantom{\dagger}}_{i+r \uparrow}
d^{\dagger}_{i \uparrow} d^{\phantom{\dagger}}_{i \downarrow} \rangle
\nonumber
\\
c^{zz'}_{\rm fd}(r) &= 
\langle f^{\dagger}_{i+r \downarrow} f^{\phantom{\dagger}}_{i+r \uparrow}
d^{\dagger}_{i \uparrow} d^{\phantom{\dagger}}_{i \downarrow} \rangle
\label{spincorr}
\end{align}
Here the superscripts $z,z'$ refer to the coordination number of the
conduction orbital on site $i$ and $i+r$ respectively.
This separation
allows us to isolate the effects of 
the number of neighbors on the spin correlations
\cite{foot1}.
We focus here on $c^{zz'}_{\rm ff}(r)$ which measures 
intersite magnetic correlations between the local electrons, 
and $c^{zz}_{\rm fd}(r=0)$, the
singlet correlator between
local and conduction electrons on the same site. 
The spin-spin correlations
are translationally invariant for uniform geometries
and periodic boundary conditions, but depend more generally on both $i$
and $r$ in irregular lattices.

We also measure the structure factor,
\begin{align}
{\cal S}^z_{\rm ff} = \sum_r \sum_{z'} c^{zz'}_{\rm ff}(r) (-1)^r
\nonumber \\
{\cal S}^{\rm tot}_{\rm ff} = \sum_{r} \sum_{zz'} g^{zz'} 
c^{zz'}_{\rm ff}(r) (-1)^r
\label{structurefactor}
\end{align}
which sums the spin-spin correlations
to all distances $r$
from sites $i$ with a given $z$.  
The staggered 
phase factor $(-1)^r$ takes the value $\pm 1$
on the two sublattices of the bipartite geometry and hence measures
AF order.
The $z$-resolved contributions to the total structure factor 
$S^{\rm tot}_{\rm ff}$
are weighted
by the fractions of sites in the lattice with a given coordination
$g^{zz'}$.  
In the singlet phase, the spin correlations decay
exponentially with separation $r$ and $S^{\rm tot}_{\rm ff}$
gets contributions only from a small number $r<\xi$ 
of local correlations.  It becomes temperature independent below
a relatively high $T$ set by the singlet
energy scale. 
In an ordered phase, on the other hand,
$S^{\rm tot}_{\rm ff}$ will depend on temperature down 
to much lower $T$ as the correlation length $\xi$ diverges.
Thus a $T$ dependence of $S_{\rm ff}^{\rm tot}$
can be used as an indicator of AF order.

Our computational approach is determinant Quantum Monte Carlo
(DQMC)\cite{blankenbecler81,hirsch85}.  This method allows the solution
of interacting tight-binding Hamiltionians like the PAM through an exact
mapping onto a problem of non-interacting particles moving in a space
and (imaginary) time dependent auxiliary field.  This field is
sampled stochastically to obtain the expectation values of different
correlation functions.  The update moves require the non-local
computation of the fermion Green's function, which also the
quantity needed to measure equal time observables
including the energy, double occupation, and spin correlations.
The algorithm involves
matrix operations and scales as the cube of the product of the number of
spatial lattice sites and the number of orbitals.  In certain special
situations, including the PAM on the geometries studied here, the
sampling is free of the sign problem\cite{loh90} so that the simulation
may be conducted on large lattices (here several hundreds of spatial
sites) at low temperature (here $T/t \lesssim 1/30$).

\section{PAM on the Lieb Lattice}

\begin{figure}
\psfig{figure=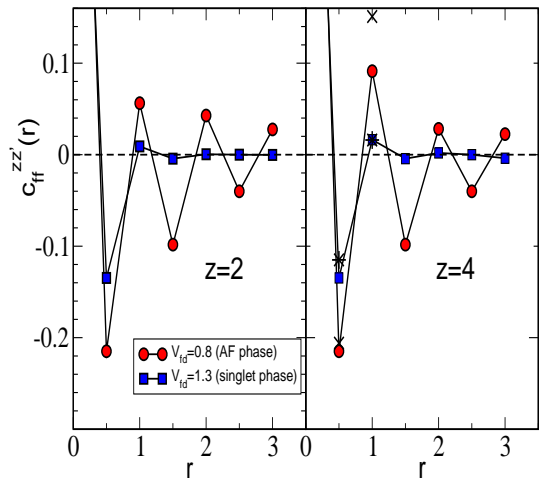,height=8.0cm,width=8.0cm,angle=0,clip} \\
\caption{(Color online) 
Spin-spin correlation function
$c^{zz'}_{\rm ff}(r)$ as a function of $r$ for 
the half-filled Lieb lattice with $U_{\rm f}=4t$ and $\beta=30$.
$r=1$ is the unit cell size: integer $r$ correspond to $z'=z$ and
half-integer values to $z' \neq z$.
The left(right) panels show $V_{\rm fd}=0.8t (1.3t)$
respectively.  For $V_{\rm fd}=0.8t$ there is AF order to large $r$,
while for $V_{\rm fd}=1.3$, 
$c^{zz'}_{\rm ff}(r)$ decays rapidly to zero.
In the AF regime, larger $z$ increases 
$c^{zz'}_{\rm ff}(r)$.
Data depicted by $x$ ($V_{\rm fd}=0.8$) and $*$ ($V_{\rm fd}=1.3$) 
are for the square lattice.  $r=0.5$ is for near-neighbors, and
$r=1.0$ for next near-neighbors.
\label{fig:corrLieb}
}
\end{figure}

We begin with the Lieb lattice which has $2N/3$ sites of coordination
number $z=2$ and $N/3$ sites with $z=4$.
Figure \ref{fig:corrLieb} shows 
$c^{zz'}_{\rm ff}(r)$
for $V_{\rm fd}=0.8$ and $V_{\rm fd}=1.3$.
In the former case, the correlation function alternates
between positive and negative values, with a correlation
length which exceeds the linear lattice size, as is
characteristic of an AF phase.  
$r=1$ corresponds to the separation between unit cells,
so that integer values of $r$ are between sites with $z'=z$ (and hence
the same sublattice) and
half-integer values have $z' \neq z$ (and hence
occupy different sublattices).
The AF correlations are evident in both $z=2$ and $z=4$, although
they are larger for higher coordination number.
This reflects the collective nature of the AF order, which is
more robust as the number of neighbors grows.
Actually, because the 
${\cal A}$ and ${\cal B}$ sublattices have different numbers,
the ordered phase is Ferrimagnetic\cite{lieb89}, with $N_{\uparrow} \neq
N_{\downarrow}$ in addition to the staggered pattern 
seen in the Figure.
For $V_{\rm fd}=1.3$, on the other hand, 
$c^{zz'}_{\rm ff}(r)$
falls rapidly to zero, indicative of a singlet phase.

\begin{figure}
\psfig{figure=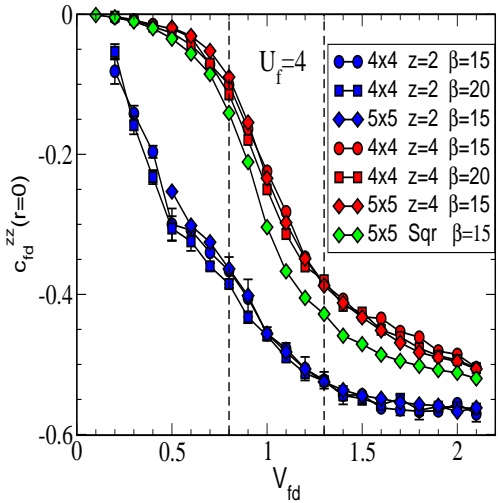,height=8.0cm,width=8.0cm,angle=0,clip} \\
\caption{(Color online) 
Local singlet correlator 
$c^{zz'}_{\rm fd}(r=0)$ for the half-filled Lieb lattice
with $U_{\rm f}=4t$.  
The singlet correlations develop more rapidly for $z=2$
than for $z=4$ since for smaller coordination number the competition
with AF order is reduced.
Data for different ($\beta=15,20$) as well as different
system sizes (4x4 and 6x6) overlap:  this short range correlation
function converges rather quickly as $T$ is lowered and $N$ is
increased.
Vertical dashed lines at $V_{fd}=0.8, 1.3$ demark the 
values used for the real space spin correlation data of Fig.~4.
Square lattice data coincide well with Lieb sites with $z=4$.
\label{fig:singletLieb}
}
\end{figure}

The AF and singlet regimes can also be distinguished by 
$c^{zz}_{\rm fd}(r=0)$, as shown in 
Fig.~\ref{fig:singletLieb}.  
(Here since $r=0$ the coordination numbers $z'=z$.)
$c^{zz}_{\rm fd}(r=0)$ vanishes for $V_{\rm fd}=0$ where the
localized and conduction fermions are decoupled, and saturates 
at a large value  for $V_{\rm fd} \rightarrow \infty$.
For the sites with larger coordination number $z'=z=4$,
singlet correlations develop at larger $V_{fd}$
than for sites with $z'=z=2$.
As might be expected for a local quantity, the singlet
correlator for the $z=4$ sites matches quite well
to those on a square lattice.  (The $4\times4$ square lattice is 
anomalous because of its unusual additional symmetries, and is
not shown.)

\begin{figure}
\psfig{figure=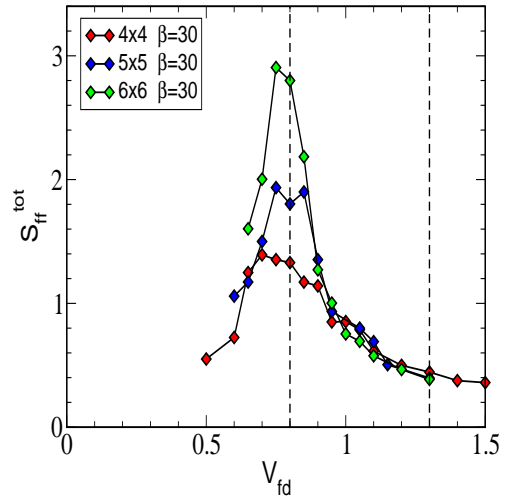,height=8.0cm,width=8.0cm,angle=0,clip} \\
\caption{(Color online) 
Localized electron antiferromagnetic structure factor for the Lieb lattice.
For $V_{\rm fd} \gtrsim 1.1$, $S_{\rm ff}^{\rm tot}$ 
is independent of temperature
and lattice size $N$.  However, when $T$ is decreased for
$V_{\rm fd} \lesssim 1.1$, $S_{\rm ff}^{\rm tot}$ grows 
as the system is cooled.
These distinct behaviors reflect the completely local nature of 
magnetic correlations in the singlet phase, and an increasing
correlation length at low $T$ in the AF phase.
Vertical dashed lines at $V_{fd}=0.8, 1.3$ demark the 
values used for the real space spin correlation data of Fig.~4.
\label{fig:SffLieb}
}
\end{figure}

In Fig.~\ref{fig:SffLieb} we turn to the AF structure factor, 
Eq.~\ref{structurefactor}, which sums the spin correlations on
the localized orbitals over the whole lattice.  In the singlet phase,
$c_{\rm ff}^{zz'}(r)$ is short ranged and temperature independent,
achieving its ground state value at $T \sim V_{\rm fd}^2/U_{\rm f}$.
In the AF phase, on the other hand, the correlation length grows as
$T$ is lowered, and hence $c_{\rm ff}^{zz'}(r)$ contributes to the 
structure factor out to larger and larger distances.
The structure factor becomes temperature dependent at low $T$.
These two regimes are evident, and are separated by $V_{\rm c} \sim 1.1.$
This is suggestive, but certainly not conclusive,
evidence of the presence of a QCP.
At the end of the following section
we will provide a finite size scaling analysis of this data to
ascertain whether there is true long range order below $V_{\rm c}$ .
Note that the reduction in $S_{\rm ff}^{\rm tot}$
as $V_{fd}$ is reduced below
$V_{fd} \approx 0.7$ is a finite temperature effect.  The RKKY exchange
scales as $V_{fd}^2$ and $T=t/30$ ($\beta t=30$) is no longer low
enough to reach the ground state.

\section{PAM on a Quasicrystal Lattice}

We turn now to the quasicrystal geometry.   Our discussion will
parallel that of the preceding section.  For this lattice,
the choices for coordination number are more numerous, $z=1,2,\cdots 8$,
as evident in Fig.~\ref{fig:quasigeom}.
The $z=1,2$ sites originate in our use of OBC,
a choice made to avoid frustration of AF order\cite{foot2}.
It is important to emphasize that these coordination numbers occur
only at the lattice edges.
Their contribution to the properties of the system
will vanish in the thermodynamic limit.

\begin{figure}
\psfig{figure=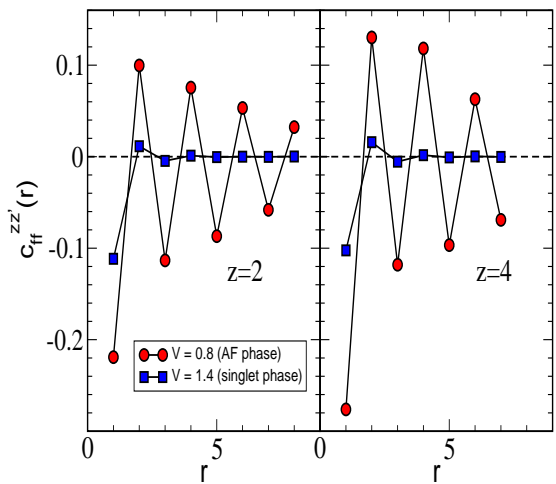,height=8.0cm,width=8.0cm,angle=0,clip} \\
\caption{(Color online) 
$z$ resolved spin-spin correlation function between localized orbitals
for $V_{\rm fd}=0.8$ (AF phase) and $V_{\rm fd}=1.4$ (singlet phase)
for the $N=41$ quasicrystal lattice at $\beta=30$.
In the former case, $c_{\rm ff}^z(r)$ remains non-zero out to large
separations, while in the latter case it falls off to zero.
Left(right) panels are $z=2(4)$.
\label{fig:corrquasi}
}
\end{figure}

$c_{\rm ff}^{zz'}(r)$ for the
quasi-crystal geometry is given in Fig.~\ref{fig:corrquasi} and shows
a differentiation between long range behavior for 
$V_{\rm fd}=0.8$ and rapid decay to zero for
$V_{\rm fd}=1.4$.
Similar to the Lieb case, 
$c_{\rm ff}^{zz'}(r)$ is larger for $z=4$ than $z=2$.
Data for other $z$ (not shown) confirm this trend.  The AF correlations
extending outward from a site become more and more robust as the
coordination number of the conduction orbital increases.

\begin{figure}
\psfig{figure=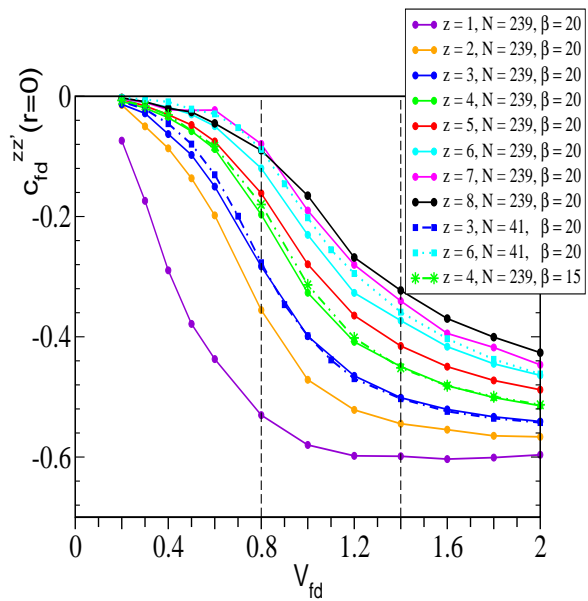,height=8.0cm,width=8.0cm,angle=0,clip} \\
\caption{(Color online) 
The singlet correlators (circles)
for the quasicrystal geometry with $N=239$ sites 
and inverse temperature $\beta=20$ shown as functions of 
$V_{\rm fd}$.  $c_{fd}^{z}(r=0)$ is largest in magnitude for smallest $z=1$.
The singlets become less and less well-formed as $z$ increases.
Data for $N=41$ sites (squares) indicate that finite size effects
are relatively small.  Similarly, data for $\beta=15$ (diamonds)
show that the low $T$ limit has been reached.
Vertical dashed lines show the $V_{\rm fd}$ values of
Fig.~\ref{fig:corrquasi}.
\label{fig:singletquasi}
}
\end{figure}

Figure \ref{fig:singletquasi} shows
the singlet correlator for the $N=239$ site quasicrystal geometry of 
Fig.~\ref{fig:quasigeom}(bottom).  The appearance of well-formed singlets
depends on the coordination number $z$ of the conduction electron site-
the point of maximum change of $c_{\rm fd}^{z}(r=0)$ shifts from 
$V_{\rm fd} \sim 0.4$ to
$V_{\rm fd} \sim 1.1$ as $z$ increases. 
This reflects the fact that AF is favored by a larger number of
neighbors, so that the cross-over to singlets requires larger $V_{\rm fd}$
as $z$ increases.
Since $c_{\rm fd}^{zz}(r=0)$ is a local quantity, its value is relatively
unaffected by total lattice size (data in
Fig.~\ref{fig:singletquasi} for $N=41$ and $N=239$ are similar),
and it also converges with $\beta$ fairly quickly.
(Data in Fig.~\ref{fig:singletquasi} for $\beta=15$ and $\beta=20$ 
are similar).

\begin{figure}
\psfig{figure=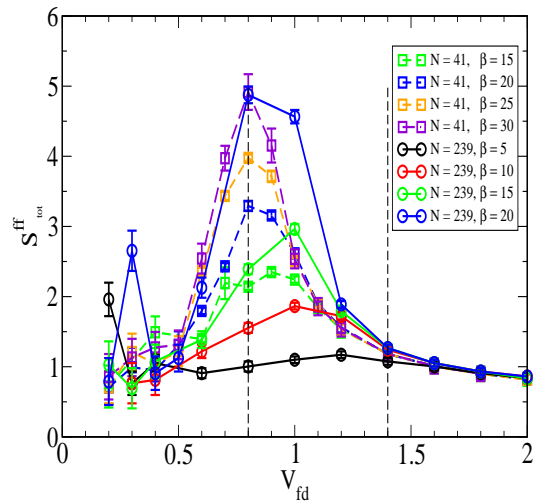,height=8.0cm,width=8.0cm,angle=0,clip} \\
\caption{(Color online) 
$S_{\rm ff}^{\rm tot}$ as a function of $V_{\rm fd}$ for 
several inverse temperatures 
$\beta$ and quasicrystal lattice sizes $N=41, 239$.  As for the
Lieb lattice, curves coincide for different $\beta$ in the singlet phase
at large $V_{\rm fd}$, but break apart at $V_{\rm fd} \approx 1.0-1.1$.
This signals the development of antiferromagnetic correlations
at large spatial separations at low $V_{\rm fd}$.
Vertical dashed lines show the $V_{\rm fd}$ values of
Fig.~\ref{fig:corrquasi}.
\label{fig:Sffquasi}
}
\end{figure}

The sum of the spin-spin correlation function of localized fermions
in the quasi-crystal geometry yields the 
structure factor and is given in Fig.~\ref{fig:Sffquasi} 
as a function of $V_{\rm fd}$.  For hybridizations
$V_{\rm fd} \gtrsim 1.1$, where results 
in Fig.~\ref{fig:singletquasi} suggest singlet formation is robust
for all $z$, $S_{\rm ff}^{\rm tot}$ is temperature independent.
Below
$V_{\rm fd} \sim 1.1$, curves
for different $\beta$ break apart, suggesting that AF correlations are
present and increasing as $T$ is lowered.  
As noted in the discussion of Fig.~6,
the reduction in the structure factor at low $V_{\rm fd}$ is 
a finite temperature effect:  the effective RKKY coupling goes as
$V_{\rm fd}^2$ and hence even larger $\beta$ is needed for AF correlations
to develop at small $V_{\rm fd}$.  See also \cite{paiva03}.

\begin{figure}
\psfig{figure=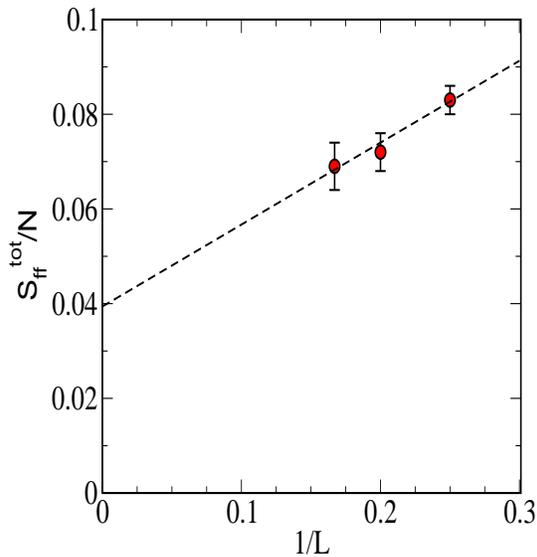,height=9.0cm,width=9.0cm,angle=0,clip} \\
\caption{(Color online) 
Finite size scaling plot for the PAM on the Lieb lattice.  Using
4x4, 6x6, and 8x8 unit cells, the
normalized structure factor $S_{\rm ff}^{\rm tot}/N$ scales to a nonzero 
value for $V_{\rm fd}=0.8$.  Here the inverse temperature $\beta=30$
which is large enough that ground state properties have been reached for
lattices of the sizes shown.
\label{fig:fss}
}
\end{figure}

In the presence of long range order (LRO) the correlation approaches
a nonzero asymptotic value $c(r\rightarrow \infty) \rightarrow m^2$,
where $m$ is the order parameter, and the structure factor
scales as $S = N m^2$.  Even if LRO is present only at $T=0$,
as is the cases in $d=2$ with continuous symmetries, this scaling
is observed at $T$ low enough that the correlation length exceeds the
largest linear lattice size studied.
We expect $S > N m^2$ on finite lattices,
since $c(r) > m^2$ at small distances, and these short range
contributions can be substantial if the lattice size is small.
A finite size scaling plot is given in Fig.~\ref{fig:fss}.

\section{Conclusions}

We have explored the competition between antiferromagnetic order
and singlet formation in the periodic Anderson model in 2D geometries
which are unfrustrated, but which have conduction electron coordination
which varies from site to site.  As is intuitively reasonable, 
singlet formation depends on $z$, and is delayed to larger
interorbital hybridization $V_{\rm fd}$ as $z$ increases.
Our data suggest that, as in the case of uniform $z$, AF order
is present in the ground state at low $V_{\rm fd}$ and absent
at large $V_{\rm fd}$.
Related issues arise in models in which site dilution
provides different conduction electron 
coordination \cite{sandvik06b,sandvik07}
or in which variation in conduction electron-local orbital
hybridization is considered \cite{castroneto98}.
It is interesting to note that the anomalous tendency for the staggered moment
to go {\it down} with increasing $z$ in the spin-1/2 Heisenberg model
on quasicrystal lattices
\cite{wessel03,wessel05,jagannathan06,jagannathan07}
would be even more evident in itinerant Hamiltonians such
as that studied here, since the greater coordination number 
reduces the local moment.  A number of experimental systems
also exhibit a similar behavior
in which $T_{Neel}$ can be higher at the (lower $z$) surface
than in the (higher $z$) bulk.\cite{zhang95}

Both geometries studied have unusual $U_{\rm f}=0$ single
particle eigenstates.  In the case of the Lieb lattice,
the inequivalence of the number of sites in the $A$ and $B$
sublattices leads to the presence of flat bands.  In the true
quasicrystal geometry (of which we explore only an approximant)
the eigenstates exhibit an intermediate scaling with
system size, between the limits where the participation
ratio ${\cal P}$ grows with the number of sites, as occurs for
Bloch states, and  ${\cal P} \sim o(1)$ for conventional localization
\cite{seo14,andrade15}.  Thus our work also explores the effects
of those features of the noninteracting spectrum on
magnetic correlations in the presence of interactions.

Limitations on accessible system sizes and temperatures prevent
us from addressing high precision questions such as whether the
distribution of $z$ destroys a sharp quantum critical point (QCP) between the 
AF and singlet phases and replaces it with a more gradual cross-over.
Related DMFT work\cite{andrade15} indicates that there is a range
of Kondo temperatures.  It is interesting to note that, to
within the accuracy DQMC simulations provide, there is remarkably
little variation between the critical value of $V_{\rm fd}$ for
different 2D conduction electron geometry.  The QCP appears to
be quite close to the square lattice value
$V_{\rm fd}/t \approx 1$ for the Lieb and
quasicrystal lattices considered here.

The usual view of the PAM is of a two band (conduction and
localized) tight binding Hamiltonian.
Although we have emphasized here the presence of sites with
different conduction electron lattice coordination numbers,
an alternate perspective on our work
is that of a study of a PAM in which the conduction 
electrons themselves have several bands.  The 
Lieb lattice geometry, for example, has three sites per unit
cell, and hence three conduction bands (Fig.~3), in addition to the
localized orbitals.  Our DQMC simulations
indicate that the competition between singlet formation
and AF order is not fundamentally affected by this more complex
band structure.

\section{Acknowledgements}

\noindent
The work of NH was supported by NSF-PHY-1263201 (REU program).
RTS and WTC were supported by DE-SC0014671.  We are grateful
to Eric Andrade and Vladimir Dobrosavljevic for useful
conversations.



\begin{thebibliography}{40}

\bibitem{rasetti91}
{\it The Hubbard Model-  Recent Results}, M. Rasetti, World Scientific,
1991.

\bibitem{montorsi92}
``{\it The Hubbard Model},'' Arianna Montorsi (ed), World Scientific,
1992.

\bibitem{gebhard97}
{\it The Mott Metal-Insulator Transition, Models and Methods},
F.~Gebhard, Springer 1997.

\bibitem{fazekas99}
{\it Lecture Notes on Electron Correlation and Magnetism}, P.~Fazekas,
World Scientific (1999).

\bibitem{white89}
``Attractive and Repulsive Pairing Vertices for the 2D Hubbard Model,"
S.R. White, D.J.~Scalapino, R.L.~Sugar, N.E.~Bickers, and
R.T.~Scalettar, Phys.~Rev.~B39, 839 (1989).

\bibitem{scalapino94}
D.J.~Scalapino, Does the Hubbard Model Have the Right Stuff?  in {\it
Proceedings of the International School of Physics} (July 1992), edited
by R.A.~Broglia and J.R.~Schrieffer (North-Holland, New York, 1994),
and references cited therein.

\bibitem{anderson61}
P.W.~Anderson, Phys.~Rev.~124, 41 (1961).

\bibitem{doniach72} 
S.~Doniach, Physica 91B, 231 (1977); B.~Cornut and B.~Coqblin, 
Phys.~Rev.~B5, 441 (1972). 

\bibitem{kondo69}
J.~Kondo, Solid State Phys.~23, 183 (1969).

\bibitem{ruderman54}
M.A.~Ruderman and C.~Kittel, Phys.~Rev.~96, 99 (1954).

\bibitem{kasuya56}
T.~Kasuya, Prog.~Theor.~Phys.~16, 45 (1956).

\bibitem{yosida57}
K.~Yosida, Phys.~Rev.~106, 893 (1957).

\bibitem{kittel63}
C.~Kittel, {\it Quantum Theory of Solids}, (Wiley, New York, 1963).

\bibitem{stewart84}
G.R. Stewart,
Rev. Mod. Phys. 56, 755 (1984).

\bibitem{stewart01}
G.R. Stewart, Rev. Mod. Phys. 73, 797 (2001).

\bibitem{gschneider78}
{\it Handboook on the Physics and Chemistry of Rare Earths},
edited by K.A. Gschneidner, Jr. and L.R. Eyring
(North Holland, Amsterdam, 1978).

\bibitem{allen82}
J.W.~Allen and R.M.~Martin,
Phys.~Rev.~Lett.~49, 1106 (1982).

\bibitem{allen92}
J.W.~Allen and L.Z.~Liu,
Phys.~Rev.~B46, 5047 (1992).

\bibitem{lavagna82}
M. Lavagna, C. Lacroix, and M. Cyrot,
Phys. Lett. 90A, 210 (1982).

\bibitem{mcmahan98}
A.~McMahan, C.~Huscroft, R.T.~Scalettar, and E.L.~Pollock,
J.~of Computer--Aided Materials Design 5, 131 (1998).

\bibitem{held01}
K.~Held, A.K.~McMahan, and R.T.~Scalettar,
Phys.~Rev.~Lett.~87, 276404 (2001).

\bibitem{blankenbecler87}
R.~Blankenbecler, J.R.~Fulco, W.~Gill, and D.J.~Scalapino, 
Phys.~Rev.~Lett.~58, 411 (1987).

\bibitem{fye90} 
R.M.~Fye, Phys. Rev. B41, 2490 (1990)

\bibitem{vekic95}
M.~Vekic, J.W.~Cannon, D.J.~Scalapino, R.T.~Scalettar, and R.L.~Sugar,
Phys.~Rev.~Lett.~74, 2367 (1995).

\bibitem{huscroft99}
C.~Huscroft, A.K.~McMahan, and R.T.~Scalettar,
Phys.~Rev.~Lett.~82, 2342 (1999).

\bibitem{loh90}
E.Y.~Loh, J.E.~Gubernatis, R.T.~Scalettar, S.R.~White,
D.J.~Scalapino, and R.L.~Sugar, Phys.~Rev.~B41, 9301 (1990).

\bibitem{troyer05}
Matthias Troyer and Uwe-Jens Wiese,
Phys. Rev. Lett. 94, 170201 (2005).

\bibitem{jarrell93}
M. Jarrell, H. Akhlaghpour, and T. Pruschke,
Phys. Rev. Lett. 70, 1670 (1993).

\bibitem{held00}
K.~Held, C.~Huscroft, R.T.~Scalettar, and A.K.~McMahan,
Phys.~Rev.~Lett.~85, 373 (2000).

\bibitem{nakatsuji02}
S. Nakatsuji, S. Yeo, L. Balicas, Z. Fisk, P. Schlottmann, P.G.
Pagliuso, N.O. Moreno, J.L. Sarrao, and J.D. Thompson,
Phys. Rev. Lett. 89, 106402 (2002).

\bibitem{ParkDropletsNature2013}
S. Seo, X. Lu, J-X. Zhu, 
R.R. Urbano, N. Curro, E.D. Bauer, V.A. Sidorov, 
L.D. Pham, T. Park, Z. Fisk, and J.D. Thompson, 
Nat. Phys. 10, 120 (2014).

\bibitem{seo14}
S. Seo, X. Lu, J-X. Zhu, R.R. Urbano, N. Curro, E.D. Bauer,
V.A. Sidorov, L.D. Pham, T. Park, Z. Fisk, and J.D. Thompson,
Nature Physics 110, 120 (2014).

\bibitem{deguchi12}
K. Deguchi, S. Matsukawa, N.K. Sato, T. Hattori, K. Ishida,
H. Takakura, and T. Ishimasa,
Nature Materials 11, 1013 (2012).


\bibitem{watanuki12}
T. Watanuki, S. Kashimoto, D. Kawana, T. Yamazaki, A.
Machida, Y. Tanaka, and T. J. Sato, Phys. Rev. B 86, 094201
(2012).


\bibitem{curro99}
N.J. Curro, Rep. Prog. Phys. 72, 026502 (2009).

\bibitem{andrade15}
E.C. Andrade, A. Jagannathan, E. Miranda, M. Vojta, and
V. Dobrosavljevic, 
Phys.~Rev.~Lett.~115, 036403 (2015).


\bibitem{grimm02}
U.W. Grimm and M. Schreiber, 
in {\it Quasicrystals - Structure and Physical Properties}, ed. H.-R.
Trebin (Wiley-VCH, Weinheim, 2003), pp. 210-235.

\bibitem{jagannathan07}
A. Jagannathan and F. Pi\'echon,
Phil.~Mag.~87, 2389 (2006).


\bibitem{goldman93}
A quasicrystal approximant is a periodic crystal whose 
atomic arrangement shares the local structure of the quasicrystal.  See 
``Quasicrystals and crystalline approximants",
A.I. Goldman and R.F. Kelton,
Rev. Mod. Phys. 65, 213 (1993).


\bibitem{wessel03}
S. Wessel, A. Jagannathan, and S. Haas, Phys. Rev. Lett.  90,
177205 (2003).

\bibitem{wessel05}
S. Wessel and I. Milat, Phys. Rev. B71, 104427 (2005).

\bibitem{jagannathan06}
A. Jagannathan, R. Moessner, and Stefan Wessel,
Phys. Rev. B 74, 184410 (2006).


\bibitem{lieb89}
E.H.~Lieb,
Phys.~Rev.~Lett.~62, 1201 (1989).

\bibitem{foot1}
One could, of course, also separate the spin correlations based on the
coordination number $z'$ of the site $i+r$.  In the case of the Lieb
lattice, where there are only two pairs of choices $z,z'=1,2$ this
would be feasible, but in the case of the quasicrystal geometry where
$z,z'=1,2,\cdots 8$ such a refined breakdown would probably do more
to obscure the analysis of the physics than to clarify it.

\bibitem{blankenbecler81}
R. Blankenbecler, D.J.~Scalapino, and R.L.~Sugar, Phys.~Rev.~D24,
2278 (1981).

\bibitem{hirsch85}
J.E.~Hirsch, Phys.~Rev.~B31, 4403 (1985).

\bibitem{foot2}
PBC are usually employed to reduce finite size effects and give
a more rapid approach to the thermodynamic limit, since they
eliminate special features like reduced coordination
number of sites at the lattice boundary.
In the quasicrystal lattice, which has a distribution of $z$
even in the absence of OBC, it is less clear that PBC are to be
preferred.  Furthermore, frustration which would be introduced
by the use of PBC would cause the formation of domain walls
which would have a similar large finite size effect as OBC.

\bibitem{paiva03}
T.~Paiva, G.~Esirgen, R.T.~Scalettar, C.~Huscroft, and A.K.~McMahan,
Phys.~Rev.~B68, 195111 (2003).

\bibitem{sandvik06b} 
A.W.~Sandvik,
Phys.~Rev.~Lett.~96 207201 (2006).

\bibitem{sandvik07} 
Kaj H.~H\"oglund, A.W.~Sandvik, and S.~Sachdev,
Phys.~Rev.~Lett.~{\bf 98}, 087203 (2007) 

\bibitem{castroneto98}
A.H.~Castro-Neto, G.~Castilla, and B.A. Jones,
Phys.~Rev.~Lett.~81, 3531 (1998).


\bibitem{zhang95}
F.~Zhang, S.~Thevuthasan,
R.T.~Scalettar, R.R.P.~Singh, and C.S.~Fadley,
Phys.~Rev.~B51, 12468 (1995).


\end{thebibliography}
\end{document}